\documentclass{PoS}

\usepackage{latexsym} 	
\usepackage{amsmath}  	
\usepackage{amsbsy}
\usepackage{epsfig}     
\usepackage{cite}

\newcommand{\be}{\begin{equation}}
\newcommand{\ee}{\end{equation}}
\newcommand{\bea}{\begin{eqnarray}}
\newcommand{\eea}{\end{eqnarray}}
\newcommand{\bean}{\begin{eqnarray*}}
	\newcommand{\eean}{\end{eqnarray*}}

\title{
Exploring phase diagram of $N_f=3$ QCD at $\mu=0$ with HISQ fermions }

\ShortTitle{Exploring phase diagram of $N_f=3$ QCD at $\mu=0$ with HISQ fermions}

\author{\speaker{H.-T. Ding}, A. Bazavov, P. Hegde, F. Karsch, S. Mukherjee and P. Petreczky
\\
Physics Department, Brookhaven National Laboratory, Upton, NY 11973\\
     E-mail: \email{htding@quark.phy.bnl.gov}
     }

\abstract{
We studied the QCD phase transition as a function of quark mass in the $N_f=3$ QCD at vanishing baryon density.
Lattice simulations have been performed using Highly Improved Staggered Quarks on $N_{\tau}=6$ lattices with quark masses that correspond
to pion masses in the region $80 \lesssim m_{\pi} \lesssim 230~$MeV. We found no evidence of the first order phase transition in the current pion mass window.
The pion mass at the critical point where the first order phase transition starts is estimated to be $m^c_{\pi}\lesssim 45$ MeV.
}

\FullConference{XXIX International Symposium on Lattice Field Theory \\
                July 10-16 2011\\
                Squaw Valley, Lake Tahoe, California}

\usepackage{epsfig}
\usepackage{graphicx}
\usepackage{dcolumn}
\usepackage{bm}

\begin{document}

\maketitle

\section{Introduction}

\begin{figure}[htp]
\begin{center}
\includegraphics[width=0.4\textwidth]{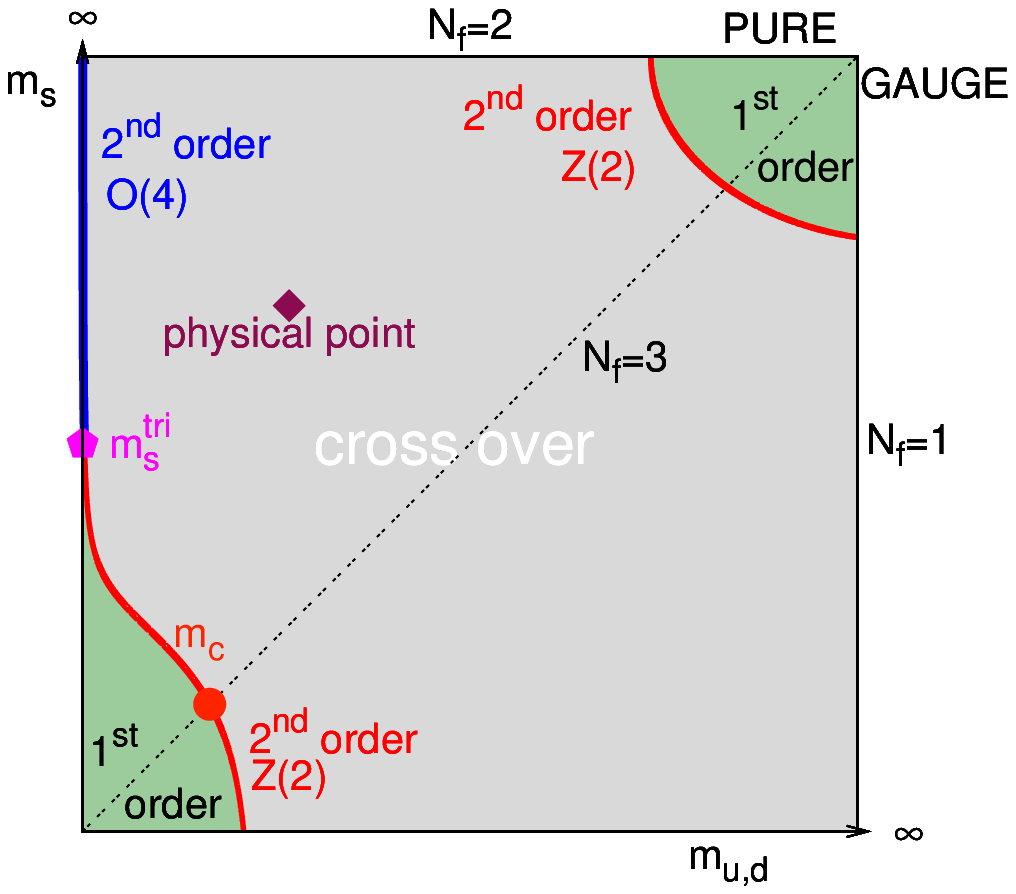}~~~
\includegraphics[width=0.5\textwidth,height=0.32\textwidth]{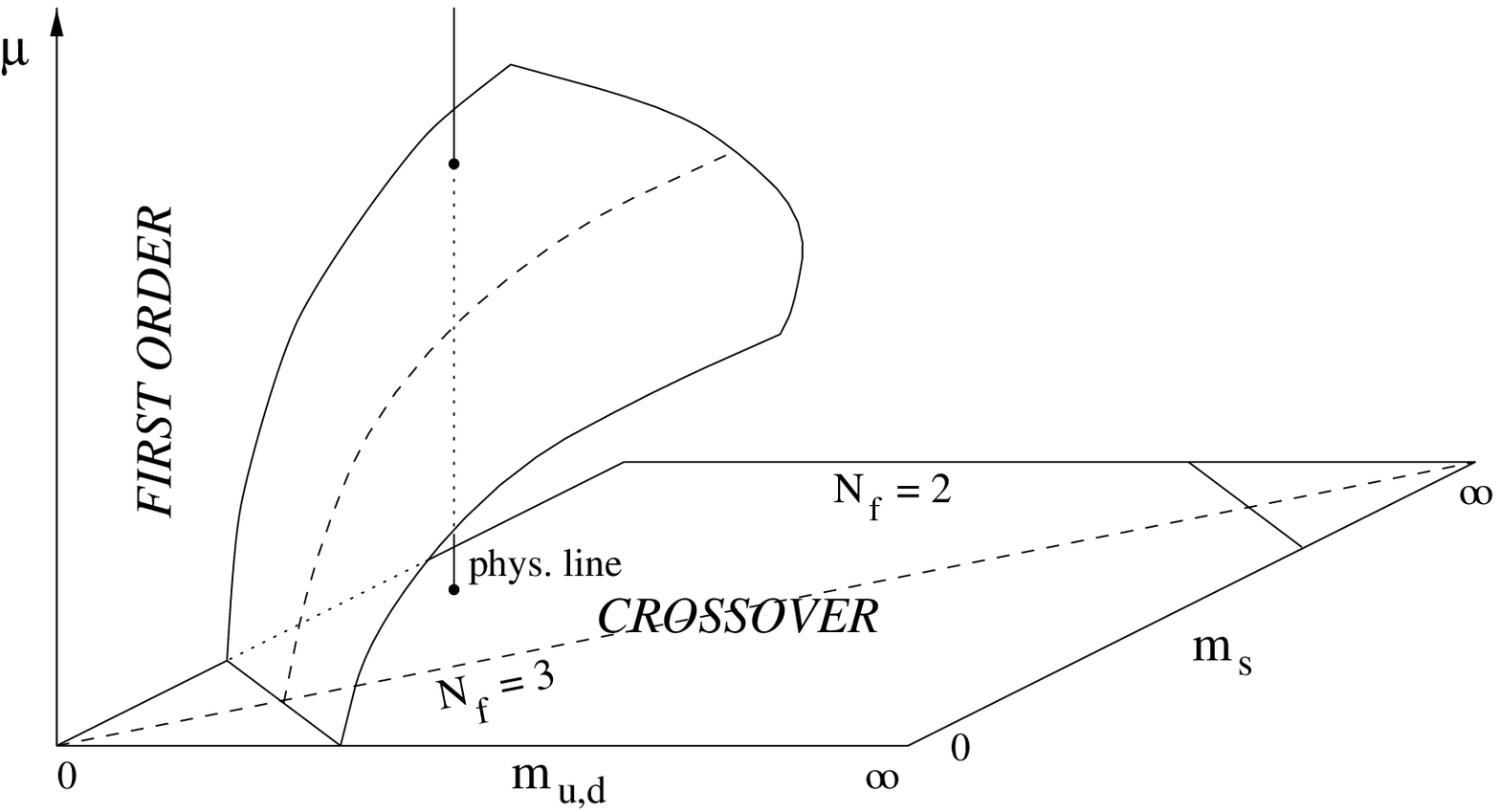}~
\end{center}
\caption{Left: schematic QCD phase transition behavior for different choices of quark 
masses ($m_{u,d}$, $m_{s}$) at zero chemical potential. Right: The critical 
surface swept by the chiral critical line at finite chemical potential. A QCD chiral critical point may exist if the surface bends towards to the physical point. The right 
plot is taken from Ref.~\cite{Karsch04}.}
\label{fig:sketch}
\end{figure}

It has been well established that at a certain high temperature and large baryon density the hadronic matter will
undergo a phase transition~\cite{KarschLecture02, reviews}. The nature of the phase transition depends on the quark 
mass and its number of flavors.
As shown in the left plot of Fig.~\ref{fig:sketch} at vanishing baryon density, in upper right corner, i.e. pure 
gauge case with infinitely heavy quark mass, there exits a first order phase transition, where the Polyakov loop 
may server as an order parameter. With decreasing 
quark mass, the first order transition will be weakened and is separated from the cross over region by 
a second order phase transition line. In the chiral limit, i.e. in the lower
 left corner of the left plot in Fig.~\ref{fig:sketch}, for 3 flavors, there exists a first order phase 
transition, whose order parameter is the chiral condensate. This first order chiral phase transition extends 
to a certain region with finite quark mass and ends at a second order phase transition line, which separates 
the cross over region from the first order phase transition region. The universal properties of the this line of the second transition are expected to be 
controlled by a global $Z(2)$ symmetry, which however is not an obvious global symmetry of the QCD Lagrangian. 
It is obvious that neither the chiral
 condensate or the Polyakov loop is an adequate order parameter for the spontaneous $Z(2)$ symmetry breaking, 
and instead the true order parameter is obtained by mixing the chiral condensate with the energy 
density~\cite{Karsch01}. When the chemical potential is turned on, the second order boundary lines turn into surfaces, as shown in the right plot of Fig.~\ref{fig:sketch}. The qualitative features of the $(T-\mu)$ phase diagram depends 
on the curvature of the surface at $\mu=0$. The common expectation is that this curvature is positive and 
there exists a chiral critical point at $\mu_c$. This expectation seems to be confirmed by lattice simulation using improved p4 action~\cite{Karsch04}. Calculations using standard staggered action on $N_{\tau}=4$ lattices on the other hard indicate a negative curvature~\cite{deForcrand:2007rq}. Thus this issue needs to be resolved in the future.

To what extent the first order phase transition holds at $\mu=0$ is of our current 
primary interests. Previous estimations using linear sigma model gives the pion mass at the critical point to be around 50 MeV ~\cite{Karsch04,Herpay:2005yr} and most recent calculations with complete one-loop parametrization of the linear sigma model predict critical pion mass to be around 110 MeV~\cite{Herpay:2006vc}. Here through lattice QCD simulations  we focus on the case with three degenerate quark masses, which corresponds to 
the dotted diagonal line shown in the left plot of Fig.~\ref{fig:sketch}, i.e. to determine the chiral 
critical point in the 3-flavor QCD. The boundary of the chiral first order phase transition in the 3-flavor 
QCD has been investigated using standard staggered actions~\cite{Christ03, deForcrand04,deForcrand:2007rq,Smith:2011pm} as well as improved staggered actions (p4, asqtad and stout)~\cite{Karsch01,Schmidt03,Karsch04, MILC05,Cheng07,Endrodi:2007gc}. There is a significant discrepancy 
of the quark masses, or the equivalently the pion masses at the chiral critical point from the above studies. Using the standard staggered action it was found that the pion mass at the critical point is about 300 MeV\cite{Christ03, deForcrand04}, while  
using improved p4 action the critical pion mass is about 70 MeV~\cite{Karsch04}. The above results are obtained on the lattices with the temporal extent $N_{\tau}=4$. Other calculations using p4 and asqtad actions on finer lattices disfavor critical pion mass of 300 MeV~\cite{Cheng07,MILC05}.
With large $N_\tau$, i.e. $N_{\tau}=6$, it has been shown that the pion mass on the critical lines becomes smaller compared 
with the one with $N_{\tau}=4$ using standard staggered fermions~\cite{deForcrand:2007rq}.  Investigations on $N_{\tau}=6$ lattices have also been performed using improved (stout) action, in which three quark flavors are not degenerate and the ratio of light to strange quark is fixed to be about 1/27 when approaching the massless limit~\cite{Endrodi:2007gc}. The analysis in Ref.~\cite{Endrodi:2007gc} suggests that the critical pion mass is about 50 MeV. Thus the first order phase transition becomes weaker when approaching the continuum limit~\footnote{ The first order phase transition also becomes weaker in the continuum limit from the study of 4-flavor QCD using HYP action~\cite{Hasenfratz:2001ef}.}.

\section{Lattice parameters}

As mentioned before,  the first order phase transition region from lattice QCD simulations shrinks with reduced cut-off 
effects. The Highly Improved Staggered Quark (HISQ) action, which is developed by the HPQCD/UKQCD collaboration~\cite{HISQ}, 
achieves better taste symmetry than the 
asqtad and p4 actions at a given lattice spacing. The improvements in HISQ action designed to 
reduce taste symmetry violations can translate into smaller lattice spacing dependences in other 
quantities and it has been found that the net results from HISQ simulations at lattice spacing $a$ 
appear to have similar lattice artifacts as that from asqtad simulations at lattice spacing
$\frac{2}{3}a$~\cite{MILC10}. Thus the HISQ simulations, which will be used in our simulations, can save substantial computing costs and can 
be essentially useful to get better understanding of the chiral first order phase transition region. Simulations have been carried out with 3 degenerate quark flavors. The quark masses vary from 0.0075 to 0.0009375 corresponding to pion masses in the region of $80 \lesssim m_{\pi} \lesssim 230~$MeV. The parameters of our simulations are given in Table~\ref{tab:parameter}.

\begin{table}[t]
\begin{center}
\vspace{0.3cm}
\begin{tabular}{|c|c|c|c|c|}
\hline
lattice dim. &   $am_q$     & $m_{\pi}$ [MeV]     &\# $\beta$ values & max. no. of traj. \\
\hline
$16^3\times$ 6           & 0.0075          & 230                       &  17                    & 6000             \\
$24^3\times$ 6          & 0.00375        & 160                    &  12                    & 7900             \\
$24^3\times$ 6           & 0.001875      & 110                      &  7                    & 8800           \\
$24^3\times$ 6          & 0.00125        &  90                       &   7                     & 5100           \\
$24^3\times$ 6           & 0.0009375    & 80                       &   8                     & 6300           \\
$16^3\times$ 6        & 0.0009375    &  80                   &   6                    & 7100          \\
\hline
\end{tabular}
\end{center}
\caption{
Parameters of the numerical simulations.
}
\label{tab:parameter}
\end{table}

\section{Universality class near critical lines}

In the vicinity of the critical lines, the free energy may be expressed as a sum of a regular and and a singular part,

\begin{equation}
f = \frac{-T}{V}\ln Z \equiv f_{\rm sing}(t,h) + f_{\rm reg}(T,m).
\end{equation}
The singular part of the free energy $f_{\rm sing}(t,h)$ is most relevant to the QCD phase transitions and dominates when system is close to the critical lines. 
The order parameter M of the transition is controlled by a scaling function that arises from the singular part of the free energy~\cite{Ejiri:2009ac,Karsch:2010ya}
\be
M(t,h)=-\partial f_{\rm sing}(t,h)/\partial h = f_{G}(z),
\ee
where $f_G(z)$ is the universal scaling function and $z=t/h^{1/\beta\delta}$. $\beta$ and $\delta$ are universal critical exponents. Here scaling variables $t$ and $h$ measure how far the system is away from the criticality. They are related to the temperature $T$ and the symmetry breaking (magnetic) field $H$,
\begin{equation}
t=\frac{1}{t_0}\frac{T-T_c}{T_c},~~~h=\frac{H}{h_0}=\frac{1}{h_0}\frac{m-m_c}{m_c},
\label{scaling_var}
\end{equation}
where $T_c$ is the transition temperature when external field $H$ vanishes, i.e. $m=m_c$. $m_c$ is the critical quark mass where transition occurs.
$t_0$ and $h_0$ are normalization factors. 

In the two flavor QCD, the order parameter $M$ for the chiral transition is the chiral condensate and the critical mass
$m_c$ is zero, thus the chiral condensate and chiral susceptibility have the following relations
\be
M= \langle\bar{\psi} \psi \rangle /T^3 \Big{|}_{\rm fixed~z}   \propto  m^{1/\delta},~~~ \chi_{q}/T^2 \Big{|}_{\rm fixed~z} \propto m^{1/\delta-1}.
\ee

In the three flavor QCD, as mentioned before, there is a first order phase transition in the chiral limit, which extends to a critical quark mass $m_c$ and ends
at a second order phase transition line. It is expected that the universal properties of this critical point are controlled by a global $Z(2)$ symmetry. 
The proper order parameter should be a mixing of two transition relevant quantities, e.g. a combination of the chiral condensate with the pure gauge action $S_G$~\cite{Karsch01}
\be
M= ( \langle\bar{\psi} \psi \rangle + r S_G) \Big{|}_{\rm T=T_c,~m_c}   \propto  (m-m_c)^{1/\delta}
\ee
and the susceptibility of the order parameter M
\be
\chi_{M}/T^2  \Big{|}_{\rm T=T_c,~m_c}  \propto  (m-m_c)^{1/\delta-1}.
\label{eq:chi_scaling}
\ee
Here $1/\delta-1$ for $Z(2)$ universal class is -0.785. With standard staggered action, it is important to construct the correct order parameter since 
$\langle\bar{\psi} \psi \rangle$ is large~\cite{Smith:2011pm}. With improved fermion action, e.g. HISQ, one may consider the effect from the mixing of the energy field is small since $\langle\bar{\psi} \psi \rangle$ is closer to zero. To make estimates on the value of $m_c$, we will use the quark chiral condensate as an approximate order parameter for the second order phase transition at the critical point.

\section{Results}
\begin{figure}[htp]
\begin{center}
\includegraphics[width=.4\textwidth]{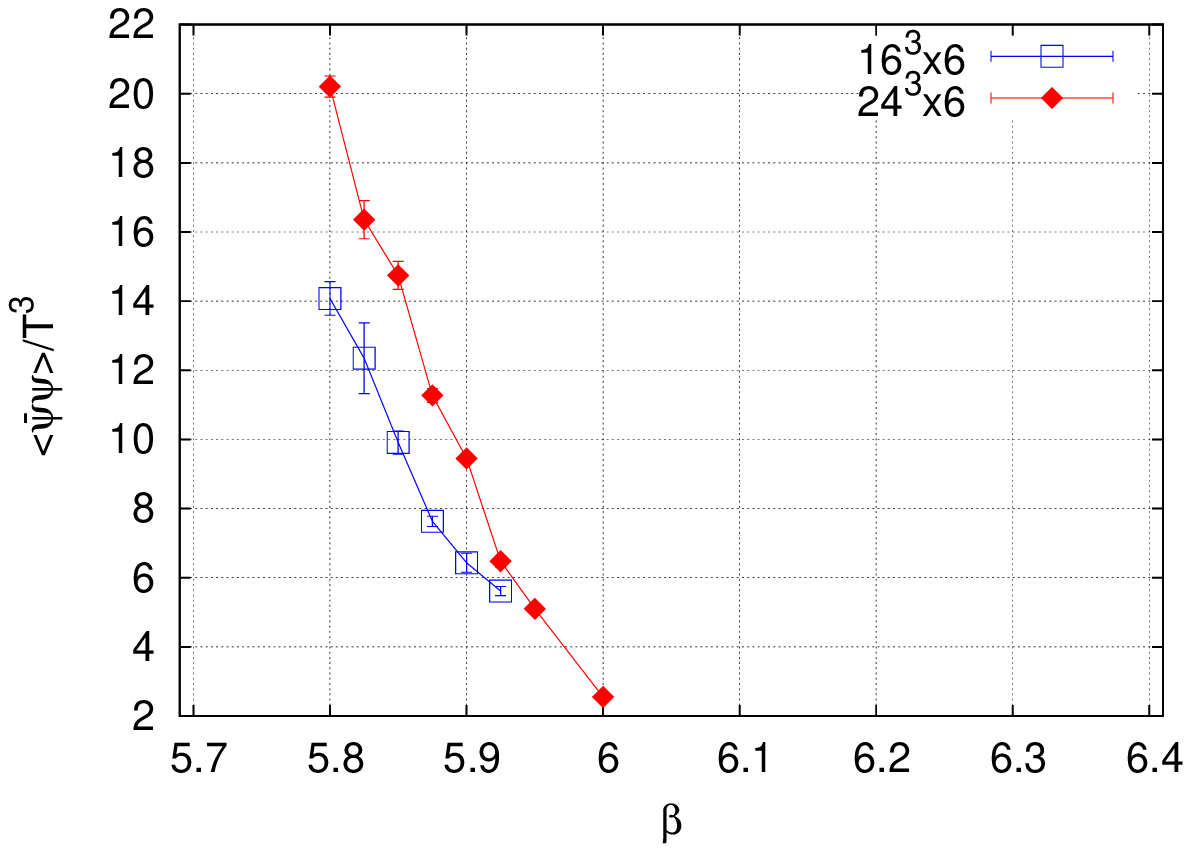}~~~\includegraphics[width=.4\textwidth]{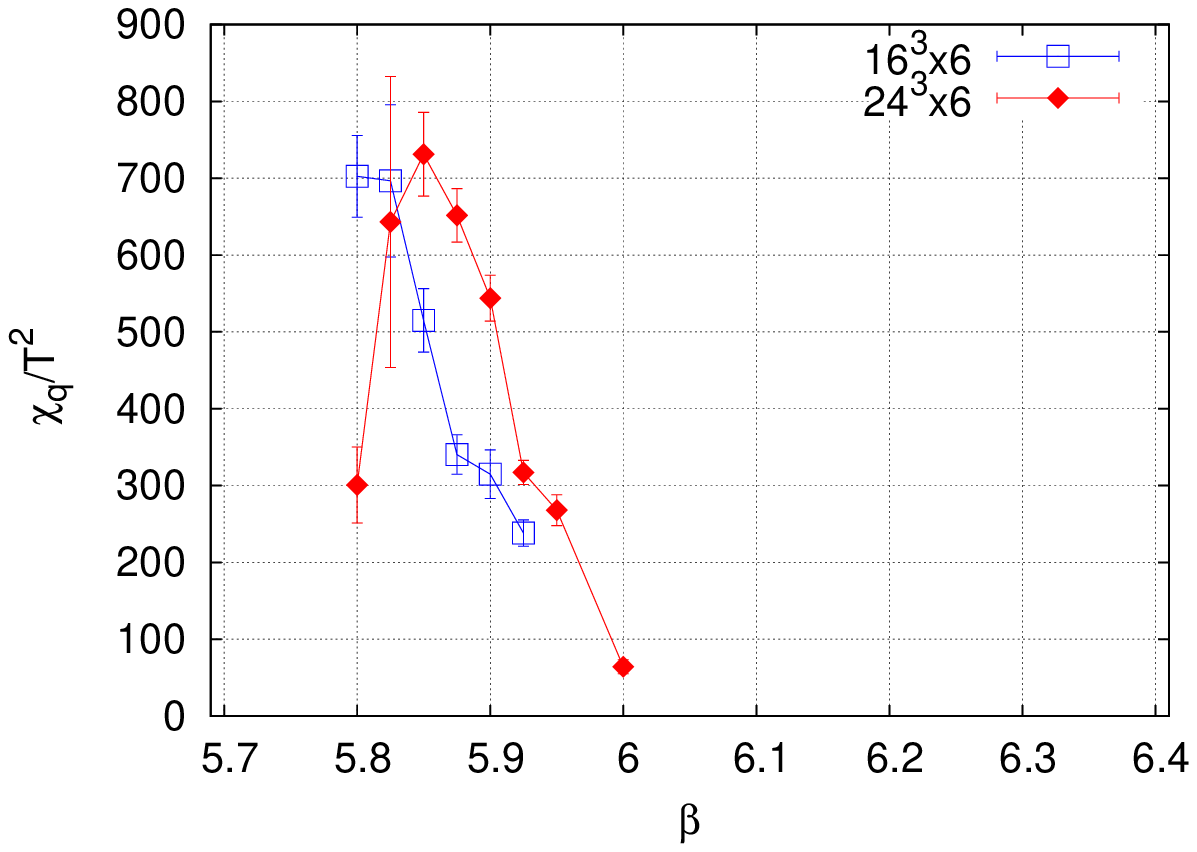}
\end{center}
\caption{Volume dependences of the chiral condensate (left) and the chiral susceptibility (right) with quark mass $am=0.0009375$.}
\label{fig:volume_dep}
\end{figure}

We first look at the volume dependence of quark chiral condensate at $am=0.0009375$ ($m_{\pi}\approx 80 $ MeV) shown in the left plot of Fig.~\ref{fig:volume_dep}.
The volume dependence at high temperature is small while at low temperature is relatively large. This is obvious since
at high temperature the scale of the system is controlled by the temperature $T$ while at low temperature the scales are the hadron masses.
In the right plot of Fig.~\ref{fig:volume_dep} we show the volume dependences of the chiral susceptibility. If the system with quark mass $am=0.0009375$
is in the first order phase transition region, the chiral susceptibility should scale with the volume. However, such volume scaling behavior is not observed with pion mass $m_{\pi}\approx80$ MeV.

\begin{figure}[htp]
\begin{center}
\includegraphics[width=.4\textwidth]{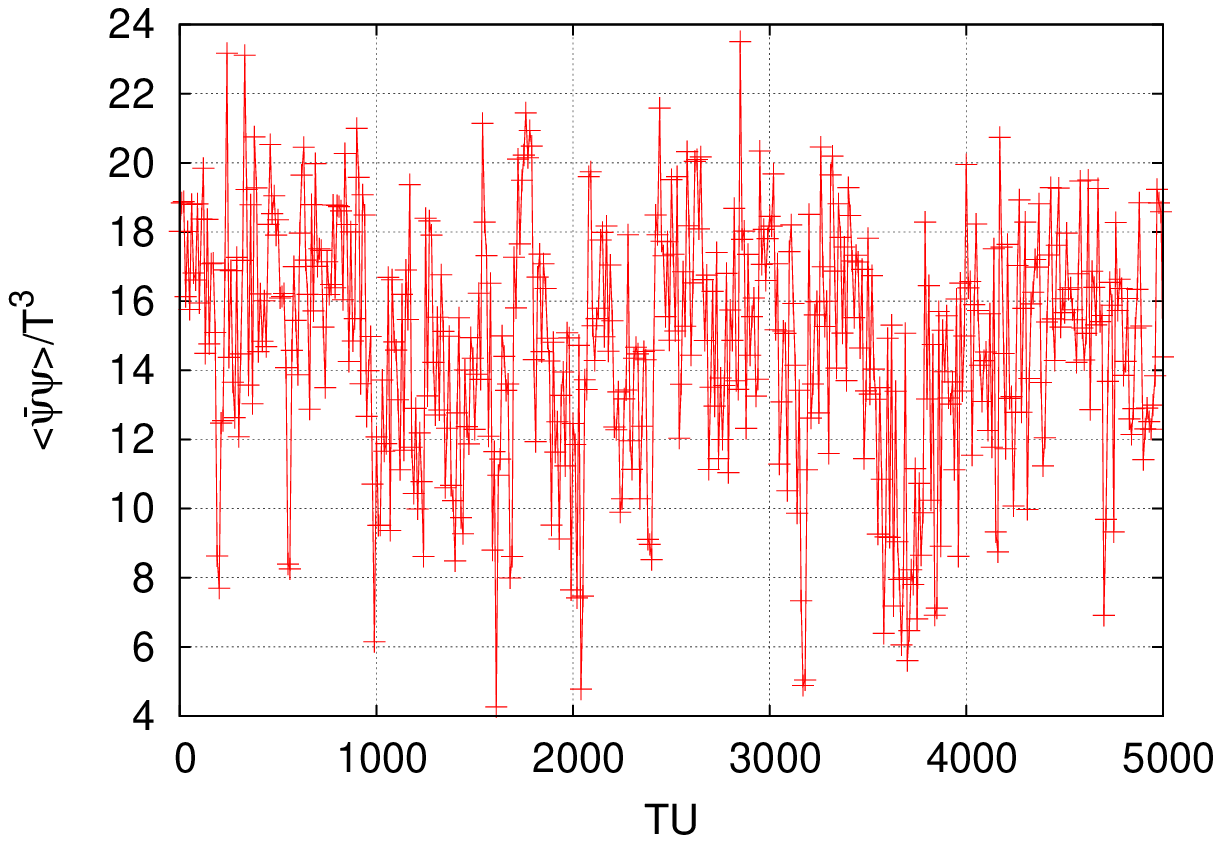}~~~\includegraphics[width=.4\textwidth]{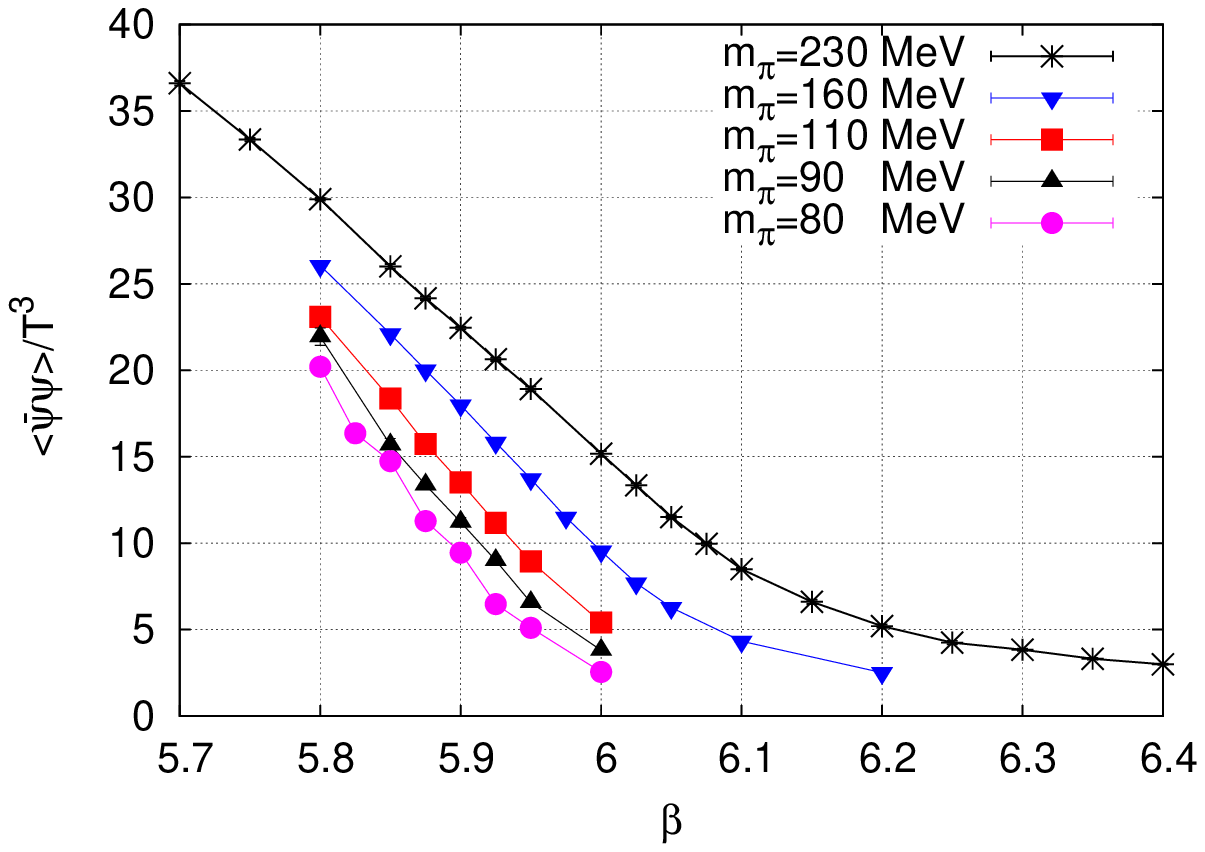}
\end{center}
\caption{Left: time history of the quark chiral condensate near $\beta_c$ with $am=0.0009375$ on $24^3\times6$ lattices. Chiral condensates as a function of $\beta$.}
\label{fig:signal}
\end{figure}

In the left plot of Fig.~\ref{fig:signal}, we show the time history of quark chiral condensate near pseudo critical beta value with our lowest quark mass on $24^3\times6$ lattices. There is no evidence for the coexistence of two phases.  We then investigate the temperature dependence of the chiral condensate at different quark masses shown in the right plot Fig.~\ref{fig:signal}. No evidence of the discontinuity of $\langle \bar{\psi}\psi\rangle$ in $\beta$ at all quark masses is found. Together with the evidence from Fig.~\ref{fig:volume_dep} we conclude that there is no first order phase transition even with quark mass down to 0.0009375 ($m_{\pi}\approx$ 80 MeV).

\begin{figure}[htp]
\begin{center}
\includegraphics[width=.4\textwidth]{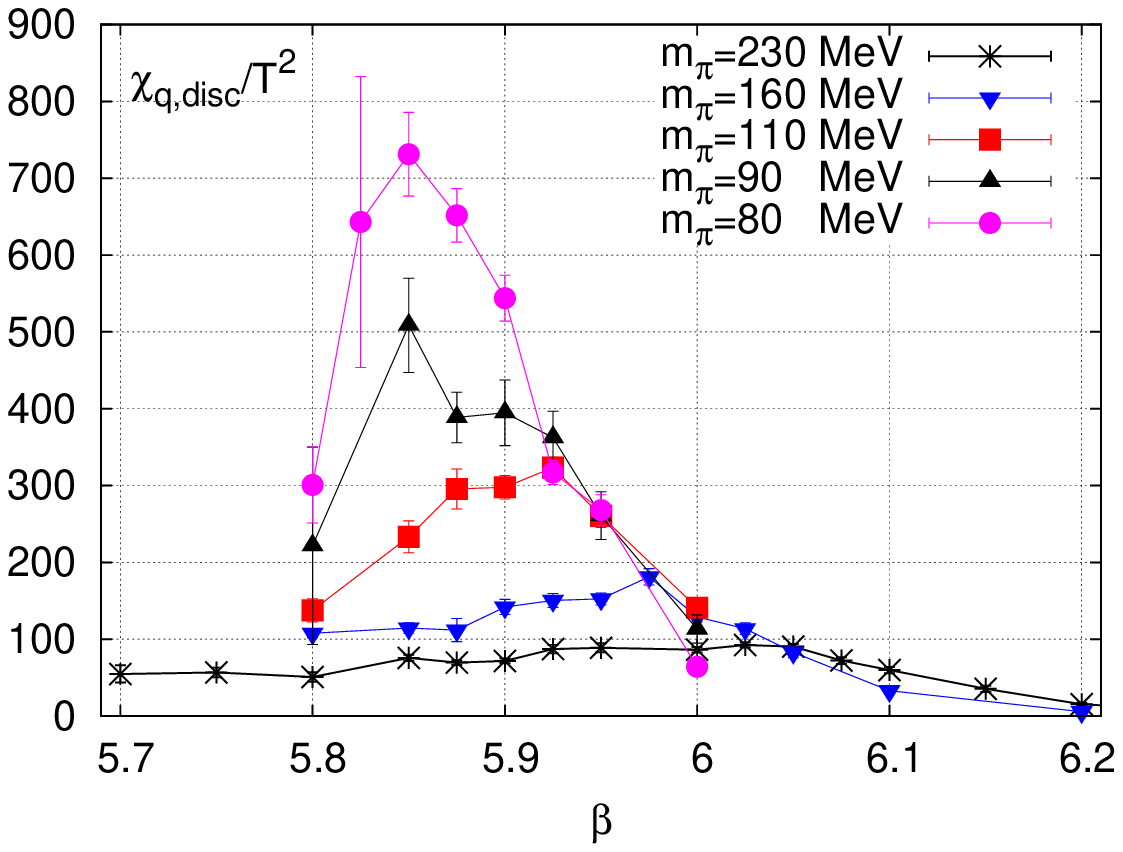}~~~~\includegraphics[width=.45\textwidth]{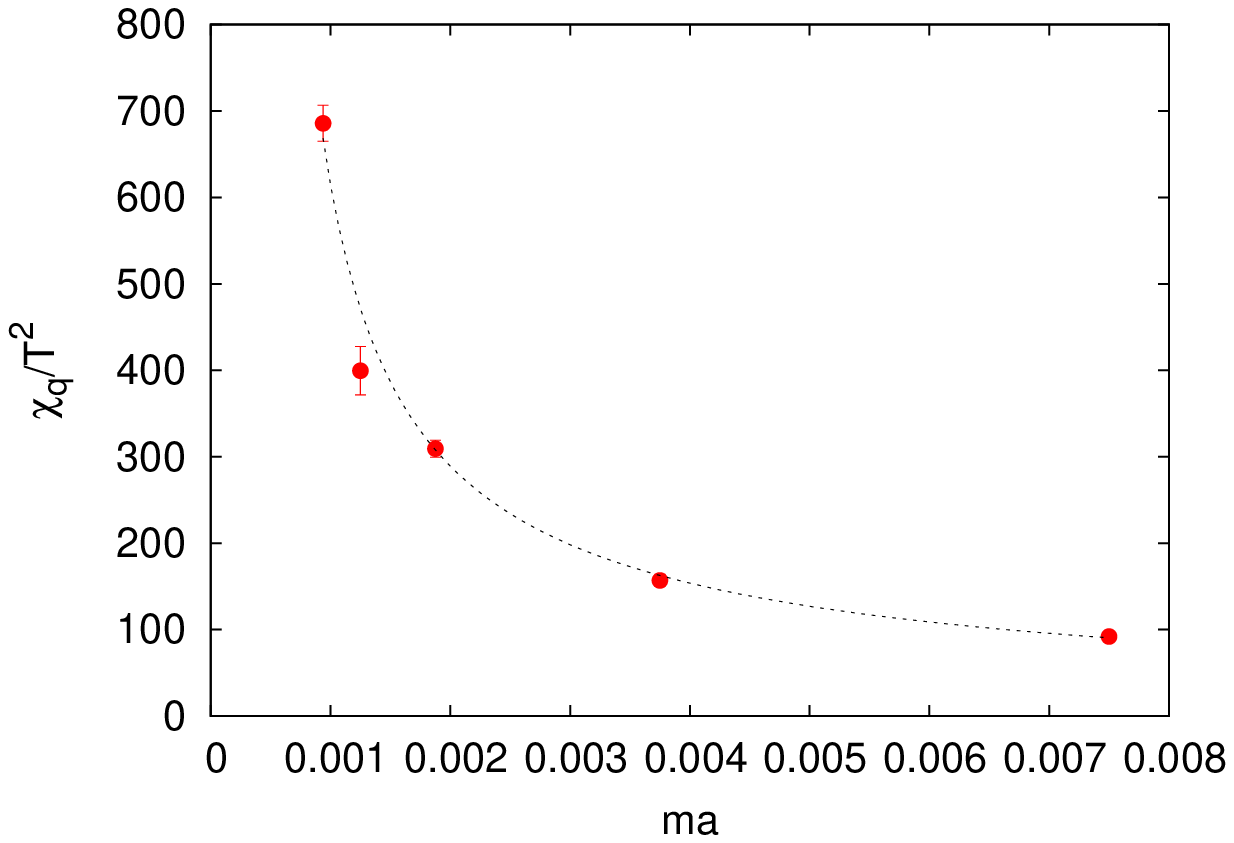}
\end{center}
\caption{Left: the disconnected part of the chiral susceptibility as a function of $\beta$. Right: the scaling fit to the height of chiral susceptibility peaks.}
\label{fig:chi_s}
\end{figure}

In the left plot of Fig.~\ref{fig:chi_s} we show the disconnected part of the chiral susceptibility as a function of quark mass. The pseudo critical temperature becomes smaller with smaller quark mass. This can be explained as that the hadronic degrees of freedom in the system become lighter and thus they become more easily excited in the thermal heat bath.  They then can contribute to the energy density of the system and thus trigger the onset of a phase transition already at a lower temperature. The peak height of chiral susceptibility grows with decreasing quark mass as the system is approaching the first order phase transition region. We then performed a scaling fit according to Eq.~(\ref{eq:chi_scaling}) to the chiral susceptibility peaks. The results are shown in the right plot of  Fig.~\ref{fig:chi_s}. As the finite volume effects would bring the peak height of chiral susceptibility up, we can get a upper bound for this analysis, which gives the pion mass at the critical point $m_\pi^{c} \lesssim 45$ MeV.

\begin{figure}[htp]
\begin{center}
\includegraphics[width=0.5\textwidth]{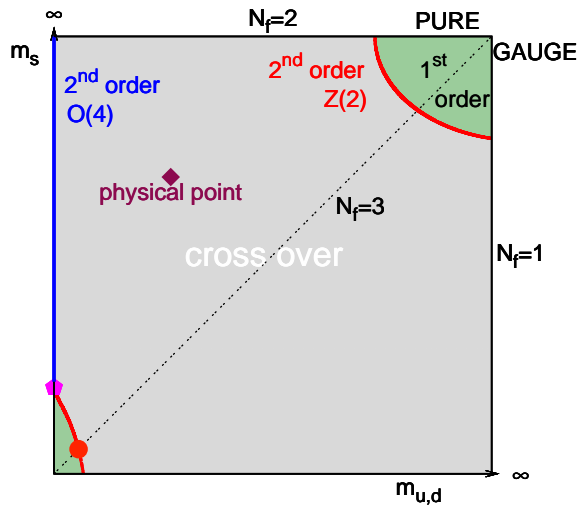}
\end{center}
\caption{QCD phase diagram at vanishing baryon density in the quark mass plane.}
\label{fig:nature}
\end{figure}

\section{Conclusion}

We have performed 3-flavor QCD simulations using HISQ/tree action on $N_{\tau}=6$ lattices with five pion masses in the region of $80\lesssim  m_\pi \lesssim~$230 MeV.
Through the study of quark chiral condensates and chiral susceptibilities, we found no evidence of the first order chiral phase transition in this pion mass region. 
The upper bound of the pion mass at the critical point is estimated to be around $45$ MeV.
It means that in the quark mass plane, the coordinate of the furthest critical point from the origin in the 3-flavor QCD is at about ($m_{\rm phy}^{s}/270,m_{\rm phy}^{s}/270$), which is very far away from the physical point at ($m^{s}_{\rm phy}/27,m^{s}_{\rm phy}$), as sketched in Fig.~\ref{fig:nature}. Together with the results from Ref.~\cite{Endrodi:2007gc}, our results suggest that the first order phase transition region is very small and thus the critical surface swept by the chiral critical line at finite chemical potential has to be bent towards to the physical point with a very large curvature to affect the nature of the real world at a small chemical potential.

\section{Acknowledgements}

The numerical simulations were carried out on clusters of
the USQCD Collaboration in Jefferson Lab and Fermilab, and on BlueGene/L computers at the New York Center for Computational Sciences (NYCCS)
at Brookhaven National Lab. This manuscript has been authored under contract number DE-AC02-98CH10886 with 
the U.S. Department of Energy.

\end{document}